\documentclass[11pt]{article}

\makeatletter\renewcommand{\section}{\@startsection
{section}{1}{\z@}{-3.5ex plus -1ex minus
    -.2ex}{2.3ex plus .2ex}{\bf }}

\makeatletter\renewcommand{\subsection}{\@startsection{subsection}{2}{\z@}{-3.25ex
plus -1ex minus
   -.2ex}{1.5ex plus .2ex}{\it }}
\makeatletter\renewcommand{\subsubsection}{\@startsection{subsubsection}{3}{-2.45ex}{-3.25ex
plus -1ex minus -.2ex}{1.5ex plus .2ex}{\it }}
\renewcommand{\thesection}{\arabic{section}.}
\renewcommand{\thesubsection}{\arabic{section}.\arabic{subsection}.}

\renewcommand{\theequation}{\thesection\arabic{equation}}
\makeatletter \@addtoreset{equation}{section}

\newcommand{\be}{\begin{equation}}
\newcommand{\ee}{\end{equation}}
\newcommand{\bea}{\begin{array}}
\newcommand{\ea}{\end{array}}
\newcommand{\beqa}{\begin{eqnarray}}
\newcommand{\eeqa}{\end{eqnarray}}

\renewenvironment{thebibliography}[1]
     {\baselineskip=16pt plus 2pt minus 1pt
      \section*{\large\refname
        \@mkboth{\MakeUppercase\refname}{\MakeUppercase\refname}}%
     \list{\@biblabel{\@arabic\c@enumiv}}%
           {\settowidth\labelwidth{\@biblabel{#1}}%
            \leftmargin\labelwidth
            \advance\leftmargin\labelsep
            \@openbib@code
            \usecounter{enumiv}%
            \let\p@enumiv\@empty
            \renewcommand\theenumiv{\@arabic\c@enumiv}}%
      \sloppy
      \clubpenalty4000
      \@clubpenalty \clubpenalty
      \widowpenalty4000%
      \sfcode`\.\@m}

\let\fn\footnote
\renewcommand{\footnote}[1]{\linespread{1.1}\fn{#1}\linespread{1.29}}

\usepackage{amsfonts}
\usepackage{amssymb}
\usepackage{mathrsfs}
\usepackage{amsmath,amssymb}
\usepackage{bbm}
\usepackage{bm}
\usepackage[vcentermath,enableskew]{youngtab}

\def\slasha#1{\setbox0=\hbox{$#1$}#1\hskip-\wd0\hbox to\wd0{\hss\sl/\/\hss}}

\def\periodb#1{\setbox0=\hbox{$#1$}#1\hskip-\wd0\hbox to\wd0{-}}

\newcommand{\unit}{\mathbbm{1}}   

\newcommand{\id}{\mathrm{id}}   
\newcommand{\CA}{\mathcal{A}}    
\newcommand{\CCA}{\mathscr{A}}    
\newcommand{\CCD}{\mathscr{D}}    
\newcommand{\CF}{\mathcal{F}}    
\newcommand{\CL}{\mathcal{L}}    
\newcommand{\CM}{\mathcal{M}}    
\newcommand{\CN}{\mathcal{N}}    
\newcommand{\CP}{\mathcal{P}}    

\newcommand{\CU}{\mathcal{U}}    
\newcommand{\frg}{\mathfrak{g}}    
\newcommand{\FK}{\mathbbm{K}}     
\newcommand{\FR}{\mathbbm{R}}     
\newcommand{\FC}{\mathbbm{C}}     
\newcommand{\CPP}{{\mathbbm{C}P}}    
\newcommand{\NN}{\mathbbm{N}}     
\newcommand{\dd}{\mathrm{d}}     
\newcommand{\dpar}{\partial}     
\newcommand{\dparb}{{\bar{\partial}}}     
\newcommand{\de}{\mathrm{e}}     
\newcommand{\di}{\mathrm{i}}     
\newcommand{\bz}{{\bar{z}}}     
\newcommand{\eps}{{\varepsilon}}     
\newcommand{\eand}{{~~~\mbox{and}~~~}}     
\newcommand{\ewith}{{~~~\mbox{with}~~~}}     
\newcommand{\der}[1]{\frac{\dpar}{\dpar #1}}   
\newcommand{\tr}{\,\mathrm{tr}\,}     

\newcommand{\au}{\mathfrak{u}}     
\newcommand{\asu}{\mathfrak{su}}     
\newcommand{\sU}{\mathsf{U}}     
\newcommand{\sSU}{\mathsf{SU}}     
\newcommand{\sG}{\mathsf{G}}     
\newcommand{\sSO}{\mathsf{SO}}     

\newcommand{\acton}{\triangleright}
\newcommand{\remark}[1]{}     

\newcommand{\appendices}{\section*{Appendix}\setcounter{subsection}{0}\setcounter{equation}{0}\renewcommand{\thesubsection}{\Alph{subsection}.}
\renewcommand{\theequation}{\thesubsection\arabic{equation}}
\makeatletter
\@addtoreset{equation}{subsection}
\makeatother
}

\def\tyng(#1){\hbox{\tiny$\yng(#1)$}}

\begin{document}
\begin{titlepage}
\begin{flushright}
  hep-th/0606197\\
  DIAS-STP-06-09
\end{flushright}
\vskip 1.0cm
\begin{center}
{\LARGE \bf Drinfeld Twist and General Relativity\\[0.2cm] with Fuzzy Spaces}
\vskip 1.5cm
{\Large Se\c{c}kin~K\"{u}rk\c{c}\"{u}o\v{g}lu \ \ and \ \ Christian S{\"a}mann}
\setcounter{footnote}{0}
\renewcommand{\thefootnote}{\arabic{thefootnote}}
\vskip 1cm
{\em School of Theoretical Physics\\
Dublin Institute for Advanced Studies\\
10 Burlington Road, Dublin 4, Ireland}\\[5mm]
{Email: {\ttfamily seckin, csamann@stp.dias.ie}} \vskip
1.1cm
\end{center}
\begin{center}
{\bf Abstract}
\end{center}
\begin{quote}
We give a simplified formula for the star product on $\CPP^n_L$, which enables us to define a twist element suited for discussing a Drinfeld twist like structure on fuzzy complex projective spaces. The existence of such a twist will have several consequences for field theories on fuzzy spaces, some of which we  discuss in the present paper. As expected, we find that the twist of the coproduct is trivial for the generators of isometries on $\CPP^n_L$. Furthermore, the twist allows us to define a covariant tensor calculus on $\CPP^n_L$ from the perspective of the standard embedding of $\CPP^n$ in flat Euclidean space. That is, we can -- in principle -- find a representation of a truncated subgroup of the diffeomorphisms on $\CPP^n$ on the algebra of functions on $\CPP^n_L$. Using this calculus, we eventually write down an Einstein-Hilbert action on the fuzzy sphere, which is invariant under twisted diffeomorphisms. 
\end{quote}
\end{titlepage}

\section{Introduction}

By now it has become rather obvious that a continuous structure of spacetime cannot persist to arbitrarily small scales. There is strong evidence from string theory that our notion of spacetime has to be enlarged to allow for supersymmetry as well as to be endowed with a noncommutative algebra of coordinates. Among the noncommutative spacetimes proposed in the literature, the so-called fuzzy spaces (\cite{Hoppe:Diss, Madore:1992av}; for a review, see \cite{Balachandran:2005ew}) play a special r{\^o}le as their algebra of functions is isomorphic to a finite-dimensional matrix algebra. Moreover, this algebra carries a representation of the full symmetry group of the corresponding commutative space, and therefore fuzzy spaces are candidates for a natural way of introducing a symmetry-preserving cut-off in quantum field theories. Scalar and gauge field theories have been discussed in considerable detail on various fuzzy spaces
\cite{Grosse:1995ar}. Fuzzy spaces also permit an elegant formulation of topologically nontrivial field configurations, such as monopoles and nonlinear sigma models \cite{Grosse:1995jt}. Supersymmetry is also treated in an exact manner in the fuzzy setting
\cite{Grosse:1995pr}.

A rather recent development in noncommutative physics has been the use of twisted Hopf algebras to recover twisted forms of those symmetries which are broken by introducing noncommutativity (see \cite{Chaichian:2004za,Wess:2003da} for Lorentz-symmetry, \cite{Matlock:2005zn} for conformal symmetries, \cite{Vassilevich:2006tc} for gauge symmetry and \cite{Kobayashi:2004ep} for supersymmetry in non-anticommutative field theories). This led eventually to the formulation of an Einstein-Hilbert action on noncommutative spaces \cite{Aschieri:2005yw,Aschieri:2005zs}, see also \cite{Balachandran:2006qg}, which is invariant under a twisted algebra of diffeomorphisms. In particular, the tensor $\theta^{\mu\nu}$ is manifestly invariant under these twisted diffeomorphisms, and thus noncommutativity of spacetime is the same for any observer. 

Until now it has not been possible to extend the Drinfeld twisting approach to fuzzy spaces majorly due to the technical difficulties presented by the rigid form of the star product on these spaces. In this article we overcome these difficulties by introducing a simplified formula for the star product of functions on $\CPP^n_L$. This enables us to define a consistent
twist element to write the twisted coproduct of symmetries acting on the algebra of functions on $\CPP^n_L$. However, the twist element corresponding to the fuzzy star product is neither unital nor does it posses a left-inverse and therefore the twist of the coproduct will destroy parts of the Hopf algebra structure. We therefore call this twist a {\em pseudo-Drinfeld twist}. 

As the algebra of isometries of $\CPP^n_L$ is not broken by fuzzification, the Drinfeld twisted approach could be deemed to be redundant in the fuzzy setting. Nevertheless, we can use it to find a representation of the group of diffeomorphisms on $\CPP^n_L$. Recent applications of the twist in gauge theory and quantum field theory provide further motivation for studying Drinfeld twisted symmetries in the fuzzy setting.

Since the algebra of functions is truncated, it is both natural and necessary to truncate also the group of diffeomorphisms to a subgroup which maps functions on $\CPP^n_L$ back to functions on $\CPP^n_L$. Furthermore, we develop a tensor calculus on $\CPP^n_L$ from the perspective of the natural embedding space $\FR^{(n+1)^2-1}\supset \CPP^n$, analogously to the usual discussion of fuzzy gauge theories. The reason for doing this is that on the (flat) embedding space, all components of tensors are described by fuzzy functions.

Having found an appropriate representation of diffeomorphisms on functions on $\CPP^n_L$, it is rather straightforward to write down a twisted-diffeomorphism invariant Einstein-Hilbert action. We perform this task in the present paper; for concreteness we specialize to $\CPP^1_L$ and give the twisted-diffeomorphism invariant Einstein-Hilbert action on this space.

Although gravity on fuzzy spaces\footnote{Another star product on $\CPP^1\cong S^2$ was given and Drinfeld twisted in \cite{Aschieri:2005zs}. This star product, however, does not correspond to the ``round'' fuzzy sphere $\CPP^1_L$.} has been discussed before in \cite{Nair:2001kr} using e.g.\ matrix models or an one-dimensional Chern-Simons action, our approach based on the twisted truncated group of diffeomorphisms is completely novel in the context of fuzzy spaces.

This paper is organized as follows. In section 2, we briefly review the geometry of fuzzy $\CPP^n$, the underlying star products as well as the isometries and their r{\^o}le in a decompactification limit. A concise review of Hopf algebras and Drinfeld twisting is given in section 3. Section 4 deals with the definition of the pseudo-Drinfeld twist and in section 5 we present the appropriate framework for discussing diffeomorphisms on $\CPP^n_L$. Eventually, the Einstein-Hilbert action on $\CPP^n_L$ is defined in section 6, where we also comment on possible physical applications and further developments.

\section{Fuzzy complex projective spaces}

\subsection{Fuzzification of $\CPP^n$}

Let us briefly recall the construction of fuzzy complex projective spaces. 
We start from coordinates $(z^\alpha,\bz^\beta)$, $\alpha,\beta=1,\ldots ,n+1$ on $\FC^{n+1}$. By imposing the normalization condition $\bz^\alpha z^\alpha=1$, we obtain a point on $S^{2n+1}\subset \FC^{n+1}$ and functions on the sphere are polynomials in these normalized coordinates. From the generalized Hopf fibration
\begin{equation}\label{generalizedHopf}
\sU(1)\ \rightarrow\  S^{2n+1}\ \rightarrow\  \CPP^n~,
\end{equation}
we obtain the functions on $\CPP^n$ as the subset of those functions on $S^{2n+1}$ which are invariant under a $\sU(1)$ action, i.e.\ the multiplication of all $z^\alpha$ by a common phase. The space of smooth functions $C^\infty(\CPP^n)$ is therefore spanned by monomials of the form
\begin{equation}\label{monomialsCPn}
z^{\alpha_1}\ldots z^{\alpha_k}\bz^{\beta_1}\ldots \bz^{\beta_k}
\end{equation}
for arbitrary $k\in\NN$. Contracting these monomials with the Gell-Mann matrices of $\sSU(n+1)$ yields  the (real) generalized spherical harmonics, which span the eigenspaces of the canonical Laplace operator on $\CPP^n$ obtained from the Fubini-Study metric.

To quantize the space, we fix the rank $k$ of the monomials \eqref{monomialsCPn} to a certain value $L$ and replace the normalized coordinates $(z^\alpha,\bz^\beta)$ by the annihilation and creation operators $(\hat{a}_\alpha,\hat{a}_\beta^\dagger)$ of $n+1$ harmonic oscillators, which satisfy the algebra $[\hat{a}_\alpha,\hat{a}_\beta^\dagger]=\delta_{\alpha\beta}$. The operators
\begin{equation}\label{genfunction}
\hat{a}^\dagger_{\alpha_1}\ldots \hat{a}^\dagger_{\alpha_L}|0\rangle\langle 0|\hat{a}_{\beta_1}\ldots \hat{a}_{\beta_L}
\end{equation}
emerging from this quantization\footnote{This way of representing functions on $\CPP^n$ is due to \cite{Dolan:2006tx}.} form an algebra of ``quantized functions'' $\hat{\CA}_L$, whose elements act on irreducible representations of $\sSU(n)$. For example, in the case $n=3$, these representations read as
\begin{equation}\label{functalgF1}
\overbrace{\overline{\tyng(4)}}^{L}~\otimes~\overbrace{\tyng(4)}^L~\
=\ ~ \overbrace{\tyng(4,4,4)}^{L}~\otimes~\overbrace{\tyng(4)}^L\ =\ ~
~\mathbf{1}\oplus\tyng(2,1,1)\oplus\tyng(4,2,2)\oplus
\ldots\oplus\overbrace{\tyng(8,4,4)}^{2L}
\end{equation}
in terms of Young diagrams. The vacua $|0\rangle\langle 0|$ in \eqref{genfunction} are inserted to simplify calculations. In principle, they could have been left out, but it will turn out to be natural to keep them. The algebra of operators closes for finite $L$, which was not the case for the monomials\footnote{Defining the product with a subsequent projection down to a certain value of $L$ yields a non-associative product structure.} \eqref{monomialsCPn}. The representation space on which the elements of $\hat{\CA}_L$ act is evidently the canonical $n+1$ oscillator Fock space. In the following, we will denote this quantized space by $\CPP^n_L$.

\subsection{Star product on $\CPP^n_L$}

As in the case of ordinary noncommutative $\FR^d_\theta$, one has a choice to work either in the operator formalism or to deform the algebra of functions one is considering by introducing a star product. Note that each point $p=(z^\alpha,\bz^\beta)$ in $\CPP^n$ corresponds to a one-dimensional vector subspace of $\FC^{n+1}$, which in turn is described by a projector $\CP(p):=(\CP(p))^{\alpha\beta}=z^\alpha \bz^\beta$ satisfying $(\CP(p))^\dagger=\CP(p)$ and $(\CP(p))^2=\CP(p)$. This projector can be expanded in terms of the Gell-Mann matrices $\lambda^a$ of $\sU(n+1)$, $\CP(p)=x^a\lambda^a$, and replacing these by the corresponding generators $\hat{\lambda}^a=\hat{a}^\dagger_\alpha\lambda^a_{\alpha\beta}\hat{a}_\beta$ in the Schwinger construction, we arrive at an operator $\hat{\CP}(p)=x^a\hat{\lambda}^a$. The $L$-fold tensor product $\hat{\rho}(p):=\hat{\CP}(p)\otimes\ldots \otimes\hat{\CP}(p)$ acts on elements of $\hat{\CA}_L$ and yields a map between elements of $\hat{\CA}_L$ and functions on $\CPP^n$ defined by
\begin{equation}\label{mapOperatorsToFunctions}
f(p)\ :=\ \tr(\hat{\rho}(p)\hat{f})~.
\end{equation}
Note that the function $f$ corresponding to an operator $\hat{f}\in\hat{\CA}_L$ via this map is a sum of homogeneous polynomials of degree $L$ in $z^\alpha$ and the same degree in $\bz^\beta$. The map \eqref{mapOperatorsToFunctions} also naturally defines a star product,
\begin{equation}
(f\star g)(p)\ :=\ \tr(\hat{\rho}(p)\hat{f}\hat{g})~,
\end{equation}
which closes in the subset of functions on $\CPP^n$ truncated at finite $L$
and therefore turns this subset into an algebra $\CA_L^\star$. Although we will focus our attention in this paper exclusively on the projector $\hat{\rho}$ given above, our choice of this projector (and therefore also the star product) is not unique; see appendix A for more details. Furthermore, the algebra of functions is not only deformed, but also truncated, contrary to the flat case. The star product is then responsible for making the truncation compatible with closure of multiplication. It is important that this star product is associative and this property is evidently inherited from the operator product on $\hat{\CA}_L$:
\begin{equation}
((f\star g) \star h)(p)\ :=\ \tr(\hat{\rho}(p)(\hat{f}\hat{g})\hat{h})\ =\ \tr(\hat{\rho}(p)\hat{f}(\hat{g}\hat{h}))=:
(f\star(g\star h))(p)~.
\end{equation}

Similarly to the star product on $\FR^d_\theta$, one can describe the one on the fuzzy space $\CPP^n_L$ by a sum of (bi-)differential operators \cite{Balachandran:2001dd}, see also the appendix:
\begin{equation}\label{defStarProduct}
(f\star g)\ :=\ \mu\left[\left(\sum_{j=0}^L\frac{(L-j)!}{L!j!}K^{\alpha_1\beta_1}\ldots K^{\alpha_j\beta_j}
\der{z^{\alpha_1}}\ldots \der{z^{\alpha_j}}\otimes\der{\bz^{\beta_1}}\ldots \der{\bz^{\beta_j}}\right)
(f\otimes g)\right]~,
\end{equation}
where $\mu(a\otimes b):=a\cdot b$ and $K^{\alpha\beta}$ is a suitable K{\"a}hler structure on $\CPP^n$, playing the r{\^o}le of the deformation tensor $\theta^{\mu\nu}$ on $\FR^d_\theta$. For our purposes, we define
\begin{equation}
K^{\alpha\beta}\ :=\ \delta^{\alpha\beta}-z^\alpha\bz^\beta~,
\end{equation}
which yields the ordinary Voros- or coherent state star product, and which is equivalent to the star product defined in \cite{Balachandran:2001dd}.

As shown in appendix B, the star product \eqref{defStarProduct} effectively reduces to 
\begin{equation}\label{StarProductReduced}
(f\star g)\ :=\ \mu\left[
\frac{1}{L!}\der{z^{\alpha_1}}\ldots \der{z^{\alpha_L}}\otimes\frac{1}{L!}\der{\bz^{\alpha_1}}\ldots \der{\bz^{\alpha_L}}
(f\otimes g)\right]~,
\end{equation}
on functions which correspond to operators in $\hat{\CA}_L$. One should stress, however, that this formula is only valid for finite $L$, and therefore the commutative limit {\em cannot} be obtained by considering the large $L$ limit of this form of the star product. Nevertheless, this formula will be particularly useful in defining the twist element later.

\subsection{Isometries on $\CPP^n_L$ and decompactification limits}

The space $\CPP^n$ can be defined as the coset space $\sSU(n+1)/\sU(n)$, and the appropriate symmetry group acting on that space and playing the r{\^o}le of the Poincar{\'e} group is therefore $\sSU(n+1)$. For each point $p\in\CPP^n$, there is a subgroup $\sU(n)$ of $\sSU(n+1)$ leaving this point invariant, and these are the rotations around a normal axis through that point. The remaining group elements correspond to translations of the point $p$.

Contrary to the star product on $\FR^4_\theta$, the star product on $\CPP^n_L$ preserves the isometries, which is one of the major advantages of fuzzy geometry compared to other regularization schemes, as e.g.\ lattice field theory.

If we consider now a decompactification limit by appropriately blowing up the neighborhood of a certain point $p\in\CPP^n_L$, we recover flat, noncommutative spacetime $\FR^{2n}_\theta$. In this limit, the translations on $\CPP^n_L$ generated by $2n$ generators of $\asu(n+1)/\au(n)$ become the $2n$ translations on $\FR^{2n}$, while the rotations $\sU(n)$ become the rotations which leave invariant the deformation tensor $\theta^{\mu\nu}$, i.e.\ the deformation tensor's stabilizer subgroup of $\sSO(2n)$.

As an example, consider $\CPP^1_L\times \CPP^1_L=\left(\sSU(2)/\sU(1)\right)_L\times\left(\sSU(2)/\sU(1)\right)_L$. Blowing up the neighborhood of the north poles of both spheres yields $\FR^4_\theta$ with the deformation tensor
\begin{equation}
\theta^{\mu\nu}\ =\ \left(\begin{array}{cccc} 0 & \theta_1 & 0 & 0\\
-\theta_1 & 0 & 0 & 0 \\
0 & 0 & 0 & \theta_2\\
0 & 0 & -\theta_2 &0
\end{array}\right)~.
\end{equation}
The two generators $J_\pm$ of both $\sSU(2)$ together become the four translations on $\FR^4$, while the two generators $J_3$ form the $\sU(1)\times \sU(1)\cong \sSO(2)\times \sSO(2)$ subgroup of $\sSO(4)$, which leaves $\theta^{\mu\nu}$ invariant and which is thus not broken by rendering $\FR^4$ noncommutative.

\section{Drinfeld twists and Hopf algebras}

Our definition of a twisted coproduct restoring twisted diffeomorphism invariance on fuzzy $\CPP^n$ will have to deviate from the usual discussion in noncommutative field theory on $\FR^n_\theta$. For this reason, let us briefly review the basics of Hopf algebras and Drinfeld twists.

\subsection{Hopf algebras}

A {\em Hopf algebra} is an algebra $H$ over a field $\FK$ together with a {\em product} $m$, a {\em unit} $\unit$, a {\em coproduct} $\Delta:H\rightarrow H\otimes H$, a {\em counit} $\eps:H\rightarrow \FK$ and an {\em antipode} $S:H\rightarrow H$. One imposes the following consistency conditions on these maps, where $g,h\in H$. The coproduct is supposed to satisfy $(\Delta\otimes\id)\Delta=(\id\otimes \Delta)\Delta$ (coassociativity) as well as $\Delta(\unit)=\unit\otimes \unit$ (unitality) and $\Delta(gh)=\Delta(g)\Delta(h)$ (homomorphism property). The counit fulfills $\eps(\unit)=1$ (unitality), $\eps(gh)=\eps(g)\eps(h)$ (homomorphism property) and $(\eps\otimes\id)\Delta=\id=(\id\otimes\eps)\Delta$ (compatibility of the product with the coproduct). The antipode (which will not play any explicit r{\^o}le in our discussion) satisfies $S(\unit)=\unit$, $m(S\otimes\id)\Delta=\eps\unit$ and $m(\id\otimes S)\Delta=\eps\unit$.

We will be exclusively interested in Hopf algebras constructed from Lie algebras. For this, consider the {\em universal enveloping algebra} $\CU(\frg)$ of a Lie algebra $\frg$, which is the tensor algebra generated by the elements of $\frg$ together with the unit $\unit$ modulo the ideal generated by the commutator relations of the algebra $\frg$.

To turn the universal enveloping algebra $\CU(\frg)$ into a Hopf algebra, we need some additional structure. On a generator $\tau$ of $\frg$, we define
\begin{equation}
\begin{aligned}
\Delta(\tau)\ =\ \tau\otimes \unit+\unit\otimes \tau&\eand \Delta(\unit)\ =\ \unit\otimes\unit~,\\
\eps(\tau)\ =\ 0&\eand \eps(\unit)\ =\ \unit~,\\
S(\tau)\ =\ -\tau&\eand S(\unit)\ =\ \unit~,
\end{aligned}
\end{equation}
and the multiplication $m$ is the algebra product $m:a\otimes b\mapsto ab$. On all other elements $\tau_1,\tau_2$ of the universal algebra $\CU(\frg)$, we define
\begin{equation}
\Delta(\tau_1\tau_2)\ =\ \Delta(\tau_1)\Delta(\tau_2)~,~~~
\eps(\tau_1\tau_2)\ =\ \eps(\tau_1)\eps(\tau_2)~,~~~
S(\tau_1\tau_2)\ =\ S(\tau_2)S(\tau_1)~.
\end{equation}
These definitions of the coproduct, the counit and the antipode have the properties demanded in the definition of a Hopf algebra, as it is easily verified.

\subsection{Review of Drinfeld twists}

The implementation of spacetime symmetries on noncommutative spaces was a 
long standing problem until very recently. It is well known that on a
$d$-dimensional noncommutative space ${\mathbb R}_\theta^d$ generated
by the coordinates $x^\mu \in {\cal A}_\theta ({\mathbb R}^d)$, the 
Poincar{\'e} and diffeomorphism symmetries are explicitly broken due to 
the noncommutativity 
\be
\lbrack x^\mu \,, x^\nu \rbrack_\star = \di \theta^{\mu \nu} \, ,
\ee
if they are naively implemented.
Very recently it has been reported in \cite{Chaichian:2004za}
and in \cite{Aschieri:2005yw} that these symmetries can be restored by twisting their coproduct. Such a twist in a general context is due to Drinfeld \cite{Drinfeld:1989st}, see also the work of Oeckl \cite{Oeckl:2000eg}. A clear way of understanding these developments is as follows \cite{Wess:2003da,Balachandran:2005eb}. 

Let $\CA$ be an algebra. For $f, g \in \CA$ there exists the multiplication map
$\mu$ such that 
\begin{equation}
\begin{aligned}
\mu : \CA \otimes \CA \ \rightarrow\  \CA~~~\\
f \otimes g \ \mapsto\  \mu (f \otimes g)~.
\end{aligned}
\end{equation}

Now let $\sG$ be the group of symmetries acting on ${\cal A}$ by a given
representation $R: \gamma \rightarrow R(\gamma)$ for $\gamma \in \sG$ and let $\frg$ be the Lie algebra of $\sG$. We can denote this action by
\be
\gamma\acton f = R(\gamma) f~.
\ee
The action of $\sG$ on ${\cal A} \otimes {\cal A}$ is formally
implemented by the coproduct $\Delta(\gamma\acton)$ and it is compatible with $\mu$ only if a certain compatibility condition between $\Delta(\gamma\acton)$ and $\mu$ is satisfied. This action is 
\be
f \otimes g \mapsto  \Delta(\gamma\acton) f\otimes g= \Delta(R(\gamma)) f \otimes g~,
\ee      
and the compatibility condition requires that 
\be
\mu \, \left (\Delta(\gamma\acton) f \otimes g \right ) = \gamma\acton \, \mu  (f \otimes g)~.
\label{eq:compatibility1}
\ee
The latter can be neatly expressed in terms of the following commutative diagram:
\begin{equation}\label{dblfibration}
\begin{aligned}
\begin{picture}(100,75)
\put(0.0,0.0){\makebox(0,0)[c]{$\mu(f\otimes g)$}}
\put(25.0,0.0){\vector(1,0){60}}
\put(120.0,0.0){\makebox(0,0)[c]{$\gamma\acton\mu(f\otimes g)$}}
\put(0.0,50.0){\vector(0,-1){40}}
\put(0.0,60.0){\makebox(0,0)[c]{$f\otimes g$}}
\put(25.0,60.0){\vector(1,0){60}}
\put(120.0,60.0){\makebox(0,0)[c]{$\Delta(\gamma\acton)f\otimes g$}}
\put(120.0,50.0){\vector(0,-1){40}}
\put(60.0,70.0){\makebox(0,0)[c]{$\Delta(\gamma\acton)$}}
\put(8.0,30.0){\makebox(0,0)[c]{$\mu$}}
\put(128.0,30.0){\makebox(0,0)[c]{$\mu$}}
\put(60.0,8.0){\makebox(0,0)[c]{$\gamma\acton$}}
\end{picture}
\end{aligned}
\end{equation}
If a coproduct $\Delta$ satisfying the above compatibility condition exists,
then $\sG$ is an automorphism of ${\cal A}$. If such a $\Delta$ 
cannot be found, then $\sG$ does not act on ${\cal A}$.   

Let us now specialize to the algebra of functions $\CA(\CM)$ on some space $\CM$ and a group of symmetries, $\sG$, with Lie algebra $\frg$. The multiplication law is the pointwise multiplication, which is deformed to a star product when $\CM$ is rendered noncommutative. Such a star product can often be defined using a twist element $\CF^{-1}\in\CU(\frg)\otimes \CU(\frg)$:
\begin{equation}
\mu(f\otimes g)\ :=\ f\cdot g ~~~\ \rightarrow\ ~~~\mu_\star(f\otimes g)\ =\ \mu(\CF^{-1} f\otimes g)~.
\end{equation}
For example, to obtain the Weyl-Moyal star product on $\FR^d_\theta$, one defines\footnote{For further examples, see \cite{Aschieri:2005zs}.}
\be
{\cal F}^{-1}_\theta\ =\ \de^{\frac{\di}{2} \theta^{\mu \nu} \partial_\mu \otimes \partial_\nu}~.
\ee
Let us denote the deformed algebra with multiplication $\mu_\star$ by $\CA^\star(\CM)$. The deformation of the product $\mu$ to $\mu_\star$ evidently requires a deformation of the coproduct, which is read off from
\begin{equation}
\xi\acton\mu_\star(f\otimes g)\ =\ \xi\acton\mu(\CF^{-1}(f\otimes g))\ =\ \mu(\Delta(\xi\acton)\CF^{-1}(f\otimes g))\ =\ \mu\left(\CF^{-1}\Delta^{\CF}(\xi\acton)(f\otimes g)\right)~,
\end{equation}
where the twisted coproduct $\Delta^\CF$ is given by
\be
\Delta^{\CF}(\xi\acton)\ =\ {\cal F} \Delta(\xi\acton) {\cal F}^{-1}\ =\ {\cal F} (\xi\acton\otimes\unit+\unit\otimes \xi\acton) {\cal F}^{-1} 
\ee 
and satisfies the compatibility condition \eqref{eq:compatibility1} by definition. 

In order to actually yield a Hopf algebra $\CU^\CF(\frg)$ with product $\mu_\star$, coproduct $\Delta^\CF$ and counit\footnote{as well as the appropriate antipode $S^\CF$, see e.g.\ the discussion in \cite{Aschieri:2005zs}} $\eps^\CF=\eps$, the twist element $\CF^{-1}$ must be an invertible, co-unital 2-cocycle. We will discuss these properties in more detail in section 4.2.

An infinitesimal symmetry $\xi\in\frg$ can be realized as a vector field $X_\xi$ acting on the algebra of functions $\CA(\CM)$. The trivial coproduct $\Delta(X_\xi)=X_\xi\otimes \unit+\unit\otimes X_\xi$ is the source of the ordinary Leibniz rule:
\begin{equation}
X_\xi\mu(f\otimes g)\ =\ \mu(\Delta(X_\xi)f\otimes g)\ =\ (X_\xi f)\cdot g+f\cdot (X_\xi g)~.
\end{equation}

The twisted coproduct will analogously give rise to a twisted or deformed Leibniz rule for the star product. It is this twisted Leibniz rule, which allows for a representation of the symmetry algebra $\frg$ on $\CA^\star(\CM)$, even after a deformation of the product $\mu$ to $\mu_\star$.

If $\CM$ is not compact, both algebras will in general be isomorphic as modules, and we denote this isomorphism by $\phi:\CA^\star(\CM)\rightarrow \CA(\CM)$. We define the action of an infinitesimal symmetry $\xi\in\frg$ on an element $f$ of $\CA^\star(\CM)$ by
\begin{equation}
\xi\acton f\ :=\ \phi^{-1}(\xi\acton\phi(f))~.
\end{equation}
On the star product of $f$ with $g\in\CA^\star(\CM)$, the action is given by the twisted coproduct:
\begin{equation}
\xi\acton(f\star g)\ =\ \mu_\star(\Delta^\CF(\xi\acton)f\otimes g)~.
\end{equation}
For further details on Drinfeld twists and Hopf algebras in this context, see \cite{Aschieri:2005zs}.

\section{Pseudo-Drinfeld twist on $\CPP^n_L$}

In this section, we define a twisted coproduct on the universal enveloping algebra $\CU(\asu(n+1))$ of the symmetry algebra $\asu(n+1)$ of $\CPP^n_L$. It is in fact not possible to define a Drinfeld twist for the star product on $\CPP^n_L$, as the appropriate twist element has no left-inverse and it is not a unital map. Therefore, twisting the coproduct will necessarily destroy parts of the Hopf algebra structure, yielding a pseudo-Drinfeld twist. However, the remaining parts will be sufficient to find a representation of the various symmetry algebras on the truncated algebra of functions $\CA_L^\star$. We will find that the coproduct of the generators of $\asu(n+1)$ remains untwisted, and thus our pseudo-Drinfeld twist of the isometries on $\CPP^n_L$ is trivial, implying that $\CU^\CF(\asu(n+1))=\CU(\asu(n+1))$.

\subsection{Definition of the twist element}

To obtain the star product on fuzzy $\CPP^n$ from a twisted product, we have to define the twist element as
\begin{equation}\label{preTwist}
\CF^{-1}_L\ :=\ \sum_{j=0}^L\frac{(L-j)!}{L!j!}\left[K^{\alpha_1\beta_1}\ldots K^{\alpha_j\beta_j}
\der{z^{\alpha_1}}\ldots \der{z^{\alpha_j}}\otimes\der{\bz^{\beta_1}}\ldots \der{\bz^{\beta_j}}\right]_{NO}~,
\end{equation}
where $[\cdot]_{NO}$ denotes a normal ordering, which puts every barred monomial into the second slot of the tensor product. For example, we have
\begin{equation}
\left[K^{\alpha\beta}\der{z^\alpha}\otimes \der{\bz^\beta}\right]_{NO}\ :=\ 
\der{z^\alpha}\otimes \der{\bz^\alpha}+z^\alpha\der{z^\alpha}\otimes \bz^\beta\der{\bz^\beta}~.
\end{equation}
Note that it is not possible to start from the star product in the projector coordinates $x$ because in these coordinates, $K^{ab}$ has terms linear in $x$, and there is no natural way of assigning the linear term to either of the slots.

As discussed in section 2.2 and appendix B, the expression \eqref{preTwist} simplifies and we define the twist element as\footnote{It is in fact this definition, and not \eqref{preTwist}, which extends to a natural but non-associative star product on the tensors on $\CPP^n_L$; see section 5.1 for more details.}
\begin{equation}
\CF^{-1}_L\ :=\ \frac{1}{L!}\der{z^{\alpha_1}}\ldots \der{z^{\alpha_L}}\otimes\frac{1}{L!}\der{\bz^{\alpha_1}}\ldots \der{\bz^{\alpha_L}}~.
\end{equation}
There are several aspects of this definition which require clarification. 
First of all, we will not discuss the twist abstractly, but always consider a representation which has as representation space the algebra of functions on $S^{2n+1}$. These functions are related to functions on $\CPP^n$ via the embedding $\FC^{n+1}\supset S^{2n+1}\supset\CPP^n$.
Second, $\CF^{-1}_L$ is actually {\em not} an element of $\CU(\asu(n+1))\otimes \CU(\asu(n+1))$, but rather an element of $\CU(\frg)\otimes \CU(\frg)$, where $\frg$ can be either considered as the Poincar{\'e} algebra $\au(n+1)\rtimes\FC^{n+1}$ of $\FC^{n+1}$ or as the algebra of diffeomorphisms (the algebra of smooth vector fields) on $S^{2n+1}$. This is, however, not a problem, as the most general objects we will encounter are functions on $S^{2n+1}$, and $\CF^{-1}_L$ will always map such functions to themselves. The multiplication operation $\mu$ then turns a product of two such functions on $S^{2n+1}$ into a function on $\CPP^n$.
And third, we denoted the twist by $\CF^{-1}_L$, following the usual nomenclature in the application of Drinfeld twists to noncommutative geometry. It is, however, rather obvious that there is no left-inverse to $\CF^{-1}_L$. As a counterexample, consider the case $L=n=1$:
\begin{equation}
\CF^{-1}_1 \big[(z^1\bz^1+z^2\bz^2)\otimes (z^1\bz^1+z^2\bz^2)\big]\ =\ \CF^{-1}_1\big[(z^1\bz^1-z^2\bz^2)\otimes(z^1\bz^1-z^2\bz^2)\big]~,
\end{equation}
and thus $\CF^{-1}_1$ is degenerate, which corresponds to the fact that e.g.\ $1\star 1=x^3\star x^3$.

Let us now show that there is, however, always a right-inverse of $\CF^{-1}_L$ on $\CA_L^\star$, where $\CA^\star_L$ is again the algebra of functions on the fuzzy space $\CPP^n_L$. For this, we will explicitly construct a right-inverse of $\der{z^\alpha}\otimes \der{\bz^\alpha}$; the right-inverse $\CF_L$ of $\CF^{-1}_L$ is then simply the $L$-fold product of the latter. The construction is done iteratively. We start from $c_0z^{\beta_1}\otimes \bar{z}^{\beta_1}$, where $c_0$ is a constant. Acting on this with $\der{z^\alpha}\otimes \der{\bz^\alpha}$ yields terms proportional to $\unit\otimes \unit$ and $z^{\beta_1}\der{z^\alpha}\otimes \bz^{\beta_1}\der{\bz^{\alpha}}$. To cancel those, one adds a term $c_1z^{\beta_1}z^{\beta_2}\der{z^{\gamma_1}}\otimes \bz^{\beta_1}\bz^{\beta_2}\der{\bz^{\gamma_1}}$. The remainder will be proportional to $z^{\beta_1}z^{\beta_2}\der{z^{\gamma_1}}\der{z^{\gamma_2}}\otimes \bz^{\beta_1}\bz^{\beta_2}\der{\bz^{\gamma_1}}\der{\bz^{\gamma_2}}$ and thus one adds the term $c_2z^{\beta_1}z^{\beta_2}z^{\beta_3}\der{z^{\gamma_1}}\der{z^{\gamma_2}}\otimes 
\bz^{\beta_1}\bz^{\beta_2}\bz^{\beta_3}\der{\bz^{\gamma_1}}\der{\bz^{\gamma_2}}$ etc.\ The total expression for the right-inverse thus reads as
\begin{equation}
\begin{aligned}
\left(\der{z^\alpha}\otimes \der{\bz^\alpha}\right)^{-1}\ =\ &c^L_0z^{\beta_1}\otimes \bar{z}^{\beta_1}+c_1z^{\beta_1}z^{\beta_2}\der{z^{\gamma_1}}\otimes \bz^{\beta_1}\bz^{\beta_2}\der{\bz^{\gamma_1}}+\\&~+c_2z^{\beta_1}z^{\beta_2}z^{\beta_3}\der{z^{\gamma_1}}\der{z^{\gamma_2}}\otimes 
\bz^{\beta_1}\bz^{\beta_2}\bz^{\beta_3}\der{\bz^{\gamma_1}}\der{\bz^{\gamma_2}}+\ldots ~.
\end{aligned}
\end{equation}
Note that this expansion will stop at terms with $L$ derivatives, and therefore convergence of this series is not a problem. The explicit values of $c_0,c_1,c_2,\ldots $ are not relevant for our further discussion. (For $L=1$ and on $\CPP^1_L$, the nontrivial coefficients are $c_0=\frac{1}{2}$, $c_1=-\frac{1}{12}$.)

Restricting the action of $\CF_L\CF_L^{-1}$ on $\CA^\star_L\otimes \CA^\star_L$, we find the formula
\begin{equation}
\CF_L\CF^{-1}_L\ =\ c_0\, z^{\alpha_1}\ldots z^{\alpha_L}\der{z^{\beta_1}}\ldots \der{z^{\beta_L}}\otimes
\bz^{\alpha_1}\ldots \bz^{\alpha_L}\der{\bz^{\beta_1}}\ldots \der{\bz^{\beta_L}}~.
\end{equation}
Note that $\CF_L\CF_L^{-1}$ is a projector.

\subsection{Consistency of the twist}

As mentioned in the introduction, the twist element will not yield a Hopf algebra after twisting the coproduct. Nevertheless, the ensuing discussion will show that the structure we obtain from the pseudo-Drinfeld twist $\CF^{-1}_L$ is sufficient for our purposes.

First of all, note that the cocycle condition 
\begin{equation}
\CF_{L;12}(\Delta\otimes \id)\CF_L\ =\ \CF_{L;23}(\id\otimes\Delta)\CF_L~,
\end{equation}
where $\CF_{L;12}=\CF_L\otimes \unit$ and $\CF_{L;23}=\unit\otimes \CF_L$, is equivalent to its inverse, even if the twist element has only a right-inverse. The inverse cocycle condition reads as 
\begin{equation}\label{inverseCocycle}
((\Delta\otimes \id)\CF_L^{-1})\CF^{-1}_{L;12}\ =\ ((\id\otimes \Delta)\CF^{-1}_L)\CF^{-1}_{L;23}~,
\end{equation}
and it is more convenient to work with as the expression for $\CF^{-1}_L$ is evidently much simpler than that for $\CF_L$. It is quite straightforward to convince oneself that this condition is indeed satisfied when acting on a product of the form $f\otimes g \otimes h$ with $f,g,h\in\CA_L^\star$. In this case, both sides of equation \eqref{inverseCocycle} read explicitly as
\begin{equation}
\dpar_{\alpha_1}\ldots \dpar_{\alpha_L}f\otimes \dparb_{\alpha_1}\ldots \dparb_{\alpha_L}\dpar_{\beta_1}\ldots \dpar_{\beta_L}g\otimes 
\dparb_{\beta_1}\ldots \dparb_{\beta_L}h~,
\end{equation}
since all derivatives which are of a higher order than $L$ in either the holomorphic or the antiholomorphic coordinates vanish trivially on $\CA_L^\star$.

The inverse cocycle condition is, as easily seen from the above discussion, related to associativity\footnote{As long as the twist element consists of {\em constant} differential operators, i.e.\ those differential operators, whose coefficients are constant functions, the inverse cocycle condition is equivalent to associativity of the star product.} of the star product, and the fact that it holds just reflects the associativity of the star product on $\CPP^n_L$. Both the cocycle and its inverse guarantee that the twisted coproduct is co-associative:
\begin{equation}
\begin{aligned}
(\Delta^\CF\otimes \id)\Delta^\CF(h)&\ =\ \big(\CF_{L;12}(\Delta\otimes \id)\CF_L\big)\big((\Delta\otimes\id)\Delta(h)\big)\big((\Delta\otimes\id)\CF^{-1}_L\CF^{-1}_{L;12}\big)~,\\
(\id\otimes\Delta^\CF)\Delta^\CF(h)&\ =\ \big(\CF_{L;23}(\id\otimes\Delta)\CF_L\big)\big((\id\otimes\Delta)\Delta(h)\big)
\big((\id\otimes\Delta)\CF^{-1}_L\CF^{-1}_{L;23}\big)
\end{aligned}
\end{equation}
for any element $h$ of the Hopf algebra. 

Neither the twist element nor its inverse are unital maps:
\begin{equation}
(\eps\otimes\id)\CF_L\neq \unit\neq (\id\otimes \eps)\CF_L~,~~~
\CF^{-1}_L(\eps\otimes\id)\neq \unit\neq \CF^{-1}_L(\id\otimes \eps)~.
\end{equation}
However, one can introduce an appropriate counit $\eps_L$, which renders the equations 
\begin{equation}\label{conditionCounit}
(\eps_L\otimes\id)\Delta^\CF(h)\ =\ h\ =\ (\id\otimes\eps_L)\Delta^\CF(h)
\end{equation}
valid. This counit is obtained by fixing
\begin{equation}\label{4.13}
\begin{aligned}
\eps_L\left(m_{\alpha_1\ldots\alpha_L}z^{\alpha_1}\ldots z^{\alpha_L}\der{z^{\alpha_1}}\ldots \der{z^{\alpha_L}}\right)&\ =\ \\ 
\eps_L\left(m_{\alpha_1\ldots\alpha_L}\bz^{\alpha_1}\ldots \bz^{\alpha_L}\der{\bz^{\alpha_1}}\ldots \der{\bz^{\alpha_L}}\right)&\ =\ 1~.
\end{aligned}
\end{equation}
The coefficients $m_{\alpha_1\ldots\alpha_L}$ in this expression count multiplicities and they are defined as
\begin{equation}
m_{\alpha_1\ldots\alpha_L}\ :=\ \left(\der{z^{\alpha_1}}\ldots\der{z^{\alpha_L}}z^{\alpha_1}\ldots z^{\alpha_L}\right)^{-1}~,
\end{equation}
where there are no sums over the indices. The definition \eqref{4.13} implies, e.g.\ that 
\begin{equation}
\eps_1\left(z^\alpha\der{z^\alpha}\right)\ =\ \eps_1\left(\bz^\alpha\der{\bz^\alpha}\right)\ =\ 1~,
\end{equation}
Wherever it is compatible with $\eps_L$ being a homomorphism, we define $\eps_L(\cdot)=0$. One can check that with the given definition of $\eps_L$, \eqref{conditionCounit} holds. Since we never need the action of $\eps_L$ explicitly, we refrain from going into details at this point. 

Summarizing, we have found that the algebra with the twisted coproduct is not as rich as usually, since the twist element is not unital and lacks a left-inverse. However, the surviving structure (a {\em bialgebra}) will prove to be sufficient for defining representations of various symmetry algebras on $\CPP^n_L$. As mentioned above, the twisted coproduct will essentially define a deformed Leibniz rule compatible with the star product and the bialgebra structure guarantees consistency of this coproduct with the ordinary product. To stress the discrepancy to the canonical twisting on $\FR^d_\theta$, we call our twist a {\em pseudo-Drinfeld twist}.

\subsection{Twist of the isometries on $\CPP^n$}

One of the reasons why Drinfeld twists on fuzzy spaces have not been considered so far might be the fact that the fuzzy version of a space has the same symmetry group as the original space, and therefore a Drinfeld twist should not yield additional symmetry. This observation amounts to the Drinfeld twist of the coproduct of the generators of $\asu(n+1)$ being trivial, which evidently continues to the coproduct of the whole enveloping algebra. Thus, the twisted coproduct should equal the untwisted one on the enveloping algebra of $\asu(n+1)$ and we briefly verify this statement by explicit calculation for our pseudo-Drinfeld twist.

The generators of $\sSU(n+1)$ which act on the truncated algebra of functions are given by
\begin{equation}
\CL^a\ :=\ \lambda^a_{\alpha\beta} z^\alpha\der{z^\beta}-\lambda^a_{\beta\alpha} \bz^\alpha\der{\bz^\beta}~,
\end{equation}
where $\lambda^a_{ij}$ are again the Gell-Mann matrices of\/\footnote{As expected, the counit acting on the generators $\CL^a$ is zero, e.g.\ for $\CPP^1$, $a=3$:
\begin{equation*}
\eps_L\big(\CL^3\big)\ =\ \eps_L\left(\sigma^3_{\alpha\beta} z^\alpha\der{z^\beta}-\sigma^3_{\beta\alpha} \bz^\alpha\der{\bz^\beta}\right)\ =\ 1-1-1+1\ =\ 0~.
\end{equation*}} $\sSU(n+1)$. They evidently preserve the degree of the monomials generating the truncated algebra of functions. We will now show that $\Delta(\CL^a)$ commutes with $\CF^{-1}_L$, and therefore $\CF^{-1}_L\Delta^\CF(\CL^a)=\CF^{-1}_L\CF_L\Delta(\CL^a)\CF^{-1}_L=\CF^{-1}_L\CF_L\CF^{-1}_L\Delta(\CL^a)=\CF^{-1}_L\Delta(\CL^a)$. 
Note that because $\CF^{-1}_L$ has only a right-inverse, we needed to include the left $\CF^{-1}_L$ stemming either from the star product $f\star g:=\mu(\CF^{-1}_Lf\otimes g)$ or another twisted coproduct $\CF_L\Delta\CF^{-1}_L$. We note that $s\Delta(\CL^a)-\Delta(\CL^a)s$ equals to
\begin{equation}\label{remainder}
\sum_{i=0}^m\left(\lambda_{\gamma_i \delta}^a\dpar_\delta\dpar_{\gamma_1}\ldots \slasha{\dpar}_{\gamma_i}\ldots \dpar_{\gamma_m}\otimes \dparb_{\gamma_1}\ldots \dparb_{\gamma_m}-\lambda_{\delta\gamma_i}^a
\dpar_{\gamma_1}\ldots \dpar_{\gamma_m}\otimes \dparb_\delta\dparb_{\gamma_1}\ldots \slasha{\dparb}_{\gamma_i}\ldots \dparb_{\gamma_m}\right)~,
\end{equation}
where $\slasha{\dpar}_{\gamma_i}$ denotes a derivative left out in the product. The expression \eqref{remainder} obviously vanishes, even for every $i$ separately. Thus, we conclude that the twisted coproduct of the enveloping algebra of the symmetry group of $\CPP^n_L$ is the same as the untwisted one, and the pseudo-Drinfeld twist construction is trivial here. 

It is important that the star commutator 
\begin{equation}
[z^\alpha\stackrel{\star}{,}\bz^\beta]\ =\ \delta^{\alpha\beta}
\end{equation}
is invariant under the action of the isometries. A straightforward calculation shows that this is indeed the case. Note also that since the coproduct is not twisted, the Hopf algebra structure of the universal enveloping algebra $\CU(\asu(n+1))$ is indeed preserved.

\subsection{Pseudo-Drinfeld twist and statistics on $\CPP^n_L$}

In the context of the Drinfeld twisted approach to Groenewold-Moyal spacetimes, it has been noticed that twisting the coproduct   
results in a deformation of the statistics of the many particle wave functions \cite{Drinfeld:1989st, Oeckl:2000eg, 
Balachandran:2005eb}. Let us first briefly recall how this comes about.

Let ${\cal A}_\theta$ be the space of single particle wave functions on the $d$-dimensional Groenewold-Moyal spacetime. Then  
the space of $n$-particle wave functions is given by the $n$-fold tensor product: ${\cal A}_\theta \otimes {\cal A}_\theta \otimes\cdots
\otimes {\cal A}_\theta$. To be more concrete let $A_1, A_2 \in {\cal A}_\theta$ and consider the two particle wave functions 
$A_1 \otimes A_2$ and $A_2 \otimes A_1$. The flip map is defined by $\sigma(A_1 \otimes A_2) = A_2 \otimes A_1$. It can
be shown that the twisted flip operator $\sigma_\theta = {\cal F}_\theta \sigma {\cal F}_\theta^{-1}$ satisfies
\begin{equation}
\lbrack \sigma_\theta \,, {\cal F}_\theta \Delta(g) {\cal F}_\theta^{-1} \rbrack \ =\  0 \eand \sigma_\theta^2 \ =\  \unit \,
\label{eq:twistedflip}
\end{equation}
where $g$ is an element of the Poincar{\'e} group. Assuming that $\sigma_\theta$ is superselected, 
we infer from (\ref{eq:twistedflip}) that the irreducible subspaces for ${\cal F}_\theta \Delta(P) {\cal F}_\theta^{-1}$ are given
by
\begin{equation}
{\cal A}_\theta^\pm \ =\  \frac{1 \pm \sigma_\theta}{2} ({\cal A}_\theta \otimes {\cal A}_\theta) \,.
\label{eq:irrsubspaces}
\end{equation}
These subspaces define the generalized bosons and fermions with the upper and the lower sign, respectively. As $\theta$ approaches zero, the usual boson and fermion statistics are recovered.

The situation is more subtle in the context of the pseudo-Drinfeld twist on $\CPP^n_L$. Here, the space of $n$-particle wave functions is denoted by ${\cal A}_L^\star \otimes 
{\cal A}_L^\star\otimes \cdots \otimes{\cal A}_L^\star$. A twisted flip operator can be defined by $\sigma_L := {\cal F}_L \sigma {\cal F}^{-1}_L$. It satisfies
\begin{equation}\label{eq:fuzzytwistedflip}
\lbrack \sigma_L \,, {\cal F}_L \Delta(g) {\cal F}_L^{-1} \rbrack \ =\  0~,~~~
\sigma^2_L\ =\ \CF_L\CF_L^{-1}~,~~~(\sigma_L^2)^2\ =\ (\sigma_L^2)~.
\end{equation}
We note that $\sigma_L$ does not square to identity as the left-inverse of ${\cal F}^{-1}_L$ does not exist. Thus, the separation of $\CA_L^\star\otimes \CA_L^\star$ into irreducible representations of $\CF_L\Delta(g)\CF_L^{-1}$ does not seem to work in quite the same way as in \eqref{eq:irrsubspaces}. A formal analogue of the split \eqref{eq:irrsubspaces} can be given as
\begin{equation}
\CF_L^{-1}\CA_L^\pm=\CF_L^{-1}\left(\frac{1\pm\sigma_L}{2}\right)\CA_L^\star \otimes \CA_L^\star~,
\end{equation}
since $\CF_L^{-1}\left(\frac{1\pm\sigma_L}{2}\right)^2=
\CF_L^{-1}\left(\frac{1\pm\sigma_L}{2}\right)$. In this equation, $\CF_L^{-1}$ can be thought of as coming from the star product, which naturally appears, since one would eventually like to compute the functional form of the two-particle wave function for the given one-particle states. We note that since $g \in \sSU(n+1)$, the twisted coproduct of the enveloping algebra of the symmetry group of $\CPP^n_L$ is the same as the untwisted one, and since we have the usual flip operator commuting with the latter, the usual statistics for bosons and fermions is present. However, if one wants to embed the isometries in the diffeomorphisms, one has to twist the coproduct of the former, as the coproduct of the latter is twisted as we will show below. We hope to report on further progress towards a better understanding of twisted statistics on $\CPP^n_L$ elsewhere. We note also that the existence of twisted statistics does not contradict the presence of the untwisted one; such a situation is already encountered on ${\mathbb R}_\theta^2$.

\section{Twist of the diffeomorphisms on the fuzzy sphere}

After defining a Drinfeld twist, let us now develop the machinery necessary for describing diffeomorphisms on the fuzzy sphere and their twisted action.  

\subsection{Tensors on fuzzy spaces}

The appropriate definition of tensors on fuzzy spaces together with a suitable associative product structure, which allows e.g.\ for Bianchi identities, has not been found\footnote{Progress in this direction will be reported in \cite{Dolan:2006tx}.} yet. Roughly speaking, it seems that components of tensors on $\CPP^n_L$ should be functions on $S^{2n+1}$, which are generated by polynomials of the form $z^{\alpha_1}\ldots z^{\alpha_L}\bz^{\beta_1}\ldots \bz^{\beta_{\bar{L}}}$ and the mismatch $L-\bar{L}$ should be related to the topological charge of the underlying bundle. The obvious generalization of the star product to a star product between the components of tensors would be the product \eqref{StarProductReduced}, which also finds a clear interpretation in terms of operator products of non-square matrices and yields the right tensor structure. However, such a product would also be non-associative and it seems not possible to extend \eqref{StarProductReduced} by derivatives of higher order than $L$ in such a way that it yields an  associative product. For this reason we will resort to embedding $\CPP^n$ into flat space $\FR^{(n+1)^2-1}$ and discuss differential geometry from the point of view of the embedding space, as it is usually done in gauge theory on fuzzy spaces. This perspective seems also natural in the light of the results presented in \cite{Grosse:2000gd}, where it was found that the exterior differential calculus on the fuzzy sphere is three-dimensional. 

The difficulties in describing bundles over fuzzy spaces are not surprising, as the latter can only be described as global objects. Differential calculus in homogeneous coordinates, however, seems to be much more subtle than simply embedding the fuzzy spaces in a flat Euclidean one.

\subsection{Lifting the discussion to the embedding space $\FR^{(n+1)^2-1}$}

For the discussion of gauge theories on the fuzzy sphere, it is necessary to introduce the concept of a gauge potential -- a Lie algebra-valued one-form -- on a fuzzy space. The usual resolution of this problem is \cite{Karabali:2001te, Balachandran:2003ay} to embed $\CPP^1$ in $\FR^3$ and use the one-forms on the embedding space. The advantage of this approach is that, since $\FR^3$ is flat, the components of all tensors are ordinary functions, and correspond in the operator formulation to square matrices. The product here is clear, and by demanding that the components of the gauge potential normal to the sphere vanish, i.e.\ 
\begin{equation}\label{ConditionOnA}
x^iA_i\ =\ 0~,
\end{equation}
one effectively reduces the theory on $\FR^3$ to $S^2\cong\CPP^1$.

One can understand this condition as coming from a reduction of the connection by
\begin{equation}
\breve{\nabla}_i\ =\ \dpar_i+\breve{A}_i~~~\ \rightarrow\ ~~~\nabla_i\ :=\ 
\di\eps_{ijk}x^j\breve{\nabla}_k\ =\ \CL_i+\di\eps_{ijk}x^j\breve{A}_k=:\CL_i+A_i~,
\end{equation}
where $\breve{A}_i$ is the unconstrained gauge potential on $\FR^3$. The condition $x^iA_i=0$ is in fact equivalent to being able to write $A_i$ as $\di\eps_{ijk}x^j\breve{A}_k$. 

In the fuzzy case, the condition $x^iA_i=0$ is no longer invariant under noncommutative gauge transformations (which is also true for $x^i\star A_i=0$) , and one has to use an alternative restriction of $A_i$ in the noncommutative case. Such a restriction was introduced in \cite{Karabali:2001te} and it reads in operator language as
\begin{equation}\label{NCcondition}
(\hat{L}^L_i+\hat{A}_i)(\hat{L}^L_i+\hat{A}_i)\ =\ L(L+1)~~~\mbox{or}~~~
\hat{A}_i\hat{L}^L_i+\hat{L}^L_i\hat{A}_i+\hat{A}_i\hat{A}_i\ =\ 0~,
\end{equation}
where $\hat{L}^L_i$ is the left action part of the adjoint action $\hat{\CL}_i=\hat{L}^L_i-\hat{L}^R_i$. This condition is indeed gauge invariant and in the large $L$ limit, together with the substitution $\frac{\hat{\CL}_i}{L}=\hat{x}_i$ it reduces to $x^iA_i=0$. To impose \eqref{NCcondition} in a field theory, one can either add it as a Lagrange multiplier to the action or turn it into a mass term, which will effectively lead to a decoupling of the normal component in numerical calculations for large masses.

In the star product formalism, the condition \eqref{NCcondition} corresponds to
\begin{equation}
x^i\star A_i+A_i\star x^i+\tfrac{1}{L}A_i\star A_i\ =\ 0~,
\end{equation}
and it is invariant under the gauge transformations
\begin{equation}
A_i\ \rightarrow\  g\star A_i\star g^{-1}+g\star \CL_i g^{-1}~.
\end{equation}

Note furthermore that the functions and the components of tensors $f$ in $\FR^3$ written as polynomials of maximal degree $L$ in $x^a/r$ automatically satisfy
\begin{equation}
r\der{r}f\ =\ x^a\der{x^a}f\ =\ 0~,
\end{equation}
where $r$ is the radial coordinate in standard spherical coordinates on $\FR^3$. Therefore, there is no need to impose this condition separately.

Although such a lift to an embedding space is in principle possible for any $\CPP^n$ (see e.g.\ \cite{Grosse:2004wm}), it is particularly simple for $\CPP^1$, as here, there is only one condition to be imposed to descend from $\FR^3$ to $S^2$. For a general $\CPP^n$, this would be the first step in reducing from $\FR^{(n+1)^2-1}$ to $S^{n^2+2n-1}$ to $\CPP^n$. Furthermore, for a physical model in four spacetime dimensions, one would like to keep time commutative and therefore the only fuzzy space which can enter in such a model is $\CPP^1_L$. 

\subsection{Geometrical structures on $\CPP^1$}

Let us briefly develop differential calculus on $\CPP^1\cong S^2$ from the perspective of the ambient space $\FR^3$. We first impose the condition
\begin{equation}\label{cond1}
r\der{r}f\ =\ x^a\der{x^a}f\ =\ 0~,
\end{equation}
on an arbitrary function $f\in C^\infty(\FR^3)$, reducing it to a function $f\in C^\infty(S^2)$. A further condition is that the metric along the radial direction should be the Euclidean one:
\begin{equation}\label{cond2}
x^ig_{ij}\ =\ x^j~.
\end{equation}
Together, \eqref{cond1} and \eqref{cond2} imply that $x^i\breve{\Gamma}_{ij}^k=0$, where $\breve{\Gamma}_k$ is the Levi-Civita connection obtained from the (unconstrained) metric $\breve{g}_{ij}$ on $\FR^3$. This is also the natural generalization of \eqref{ConditionOnA}. We thus proceed as in the case of gauge theories and define
\begin{equation}
\nabla_i\ :=\ \di\sqrt{|g|}\eps_{ijk}x^j g^{kn}\dpar_n+\di\sqrt{|g|}\eps_{ijk}x^j\breve{\Gamma}_{n}g^{kn}\ =:\ \CL_i+\Gamma_i~.
\end{equation}
Note that the additional factor of $\sqrt{|g|}$ guarantees that $\nabla_i$ indeed transforms as a one-form under diffeomorphisms. The curvature tensor is then naturally given by
\begin{equation}
\begin{aligned}
R^k{}_{lij}\ :=\ &([\nabla_i,\nabla_j])_l^k-\di\sqrt{|g|}\eps_{ijm}g^{mn}(\nabla_n)_l^k\\
\ =\ &\CL_i\Gamma^k_{jl}-\CL_j\Gamma^k_{il}+\Gamma^k_{in}\Gamma^n_{jl}-\Gamma^k_{jn}\Gamma^n_{il}
-\di\sqrt{|g|}\eps_{ijm}\Gamma_{nl}^k g^{mn}~.
\end{aligned}
\end{equation}
From here, we can follow the ordinary discussion and introduce the Ricci tensor and the curvature scalar by
\begin{equation}
R_{mn}\ :=\ R^i{}_{min}\eand R\ :=\ R_{mn}g^{mn}~.
\end{equation}
The Einstein-Hilbert action then simply reads as
\begin{equation}
S=\int\dd^3 x \sqrt{|g|} R \delta(x^a x^a-1)
\end{equation}
and is evidently invariant under diffeomorphisms which have a trivial radial part.

It will also be useful to have at hand an alternative approach to describe diffeomorphisms on the sphere $S^2$. Such a formulation is found by considering the spherical harmonics as elements of a basis $e=(e^i)$ of the infinite dimensional vector space of smooth functions. We can then define a (finite) general coordinate transformation $D_M\in\CCD:x\rightarrow f(x)$ as an invertible linear map $M=(M^m{}_i)$ acting according to
\begin{equation}\label{finitediffeos}
x^m\ \rightarrow\  \tilde{x}^m\ :=\ D_M\acton x\ :=\ M^m{}_i e^i~,
\end{equation}
where the $M^m{}_i$ are real, $m=1,...,3$ and $i\in\NN$. By the usual statement that a diffeomorphism induces a pullback of the coordinates, which can be understood as a general coordinate transformation, we can in fact describe all diffeomorphisms in this way. 

This idea has been used e.g.\ in \cite{Sasakura:2005js}. It is quite obvious that one can easily truncate the group of diffeomorphisms to an appropriate subgroup, which has $\CA_L^\star$ in a natural manner as its representation space. 

\subsection{Diffeomorphisms on $\CPP^1_L$}

As the functions $\CA_L^\star$ on the fuzzy space $\CPP^1_L$ are a subset of the full algebra of functions, it is evident that also the diffeomorphisms, which are maps $\CPP^1_L\rightarrow \CPP^1_L$, have to be a subset of those on $\CPP^1$. In the case of the fuzzy sphere, the number of spherical harmonics is finite, and therefore a fuzzy diffeomorphism $D_M^L\in \CCD_L$ is simply given by a finite dimensional matrix $M$ according to
\begin{equation}\label{finitefuzzydiffeos}
x^m\ \rightarrow\  \tilde{x}^m\ :=\ D_M^L\acton x\ :=\ M^m{}_i e^i~,
\end{equation}
where $m=1,...,3$ and $i=1,...,(L+1)^2$. After imposing the condition that we only admit those $D_M^L$, whose action $M$ is invertible, the fuzzy diffeomorphisms form evidently a group. Invertibility means here that there is an $\tilde{M}=(\tilde{M}^m{}_i)$, which acts on the basis $\tilde{e}^i$ such that $\tilde{M}^m{}_i\tilde{e}^i=x^m$. This group is furthermore a subgroup of the group of diffeomorphisms on $\CPP^1$, as one can embed the truncated $M^\mu{}_i$ into an infinite dimensional matrix by adding columns of zeros to its right side. We denote this embedding by $\phi_\CCD:\CCD_L\rightarrow\CCD$.

Note that one can write the action of any $D_M^L$ on $\CA_L^\star$ in terms of a differential operator of order $L$ while for the diffeomorphisms on $\CPP^1$, one needs generally an infinite series of differential operators of arbitrarily high order. In terms of complex coordinates, the diffeomorphisms are multidifferential operators of the form
\begin{equation}
D_M^L\acton\ =\  \sum_{I,J,K,L} c_{IJ}^{KL}z^I\bz^J\der{z^K}\der{\bz^L}~,
\end{equation}
where $I,J,K,L,$ are multi-indices with $|I|=|K|$ and $|J|=|L|$. In terms of the real coordinates, we have 
\begin{equation}
D_M^L\acton\ =\  \sum_{I,J} d_{I}^{J}x^I\der{x^J}~,
\end{equation}
where $I$ and $J$ are multi-indices with $|I|<|J|$. It is rather obvious that this definition of fuzzy diffeomorphisms easily generalizes to all fuzzy complex projective spaces.

The group $\CCD_L$ is clearly non-trivial, as it contains e.g.\ the isometry group of $\CPP^1$, $\sSU(2)\cong\sSO(3)$. Its generators are given by 
\begin{equation}
\CL^a\ =\ \sigma^a_{\alpha\beta}z^\alpha\der{z^\beta}-\sigma^a_{\alpha\beta}\bz^\alpha\der{\bz^\beta}~,
\end{equation}
where $\sigma^a$ are the Pauli matrices, or, in the projector coordinates $x^a$, by
\begin{equation}
\CL^a\ =\ \tau^a_{bc}x^b\der{x^c}~,
\end{equation}
where $\tau^a_{bc}$ are the Gell-Mann matrices of $\sSO(3)$. In the basis $e$ consisting of $1,x^1,x^2,x^3,\ldots $, the matrices $M$ representing isometries are block diagonal matrices of the form
\begin{equation}
M\ :=\ \left(\begin{array}{cccc} 0 & & 0 & \ldots   \\ 0 &  M_3 & 0 & \ldots  \\0 & & 0 & \ldots 
\end{array}\right)\ewith M_3\in\sSO(3)~.
\end{equation}

Let us now introduce an extension of the algebra of functions $\CA$ to the algebra of tensors and not necessarily covariant derivatives of tensors $\CCA$. Note that there is an action of diffeomorphisms on $\CCA$. Truncating the Taylor expansions of the components of these objects at monomials of degree $L$ in $x^a$, we obtain the corresponding algebra $\CCA^\star_L$, which is naturally embedded in $\CCA$ by the map
\begin{equation}
\phi_\CCA:\CCA^\star_L\ \rightarrow\  \CCA
\end{equation}

We define the action of a fuzzy diffeomorphism on an element of $\CCA^\star_L$ with the help of the embeddings $\phi_\CCD$ and $\phi_\CCA$. That is, the action of a diffeomorphism $D_M^L$ on an element $A_L\in\CCA_L^\star$ is defined as
\begin{equation}
D_M^L\acton A_L\ :=\ \phi^{-1}_\CCA\left(\phi_\CCD(D^L_M)\acton\phi_\CCA(A_L)\right)~,
\end{equation}
where $\phi_\CCD(D^L_M)\acton\phi_\CCA(A_L)$ is the ordinary action of a diffeomorphism in $\FR^3$ on objects in $\CCA$ and the restriction of $\CCD$ to $\CCD_L$ guarantees the existence of the inverse $\phi^{-1}_\CCA$ of the result.

The action of a diffeomorphism $D_M\in\CCD$ on a product of $A_1,A_2$ of $\CCA$ is defined via the coproduct of $D_M$:
\begin{equation}
D_M\acton (A_1\cdot A_2)\ :=\ \mu(\Delta(D_M\acton) A_1\otimes A_2)
\end{equation} 
To lift this representation of diffeomorphisms to the deformed algebra $\CA_L^\star$, we have to twist this coproduct and obtain
\begin{equation}
D_M^L\acton (A_1\star A_2)\ =\ \mu(\Delta(D_M^L\acton)\CF^{-1}_LA_1 \otimes A_2)\ =\ \mu(\CF^{-1}_L\Delta^\CF(D_M^L\acton) A_1\otimes A_2)~.
\end{equation}

Furthermore, it is important to verify that the star product of two tensors transforms correctly to make sure that we have a consistent representation of the truncated algebra of diffeomorphisms. The proof for this is essentially the same as in the case of the Weyl-Moyal star product on $\FR^4_\theta$ and discussed in \cite{Aschieri:2005yw}. It is, however, constructive to look at the proof that  the star product of two functions transforms indeed as a function e.g.\ in the case $L=1$. Let us consider an infinitesimal diffeomorphism ${\hat \delta}_D$. We thus have
\begin{equation*}
\begin{aligned}
\mu(\Delta({\hat \delta}_D ) &\CF^{-1}_L A_1 \otimes A_2)\\
 &\ =\  \mu \left ( ( {\hat \delta}_D  \otimes 1 + 1 \otimes 
{\hat \delta}_D  ) \delta^{\alpha \beta} \partial_\alpha A_1 \otimes {\bar \partial}_\beta A_2 \right ) \\
&\ =\  \mu \left({\hat \delta}_D (\partial_\alpha A_1 \delta^{\alpha \beta}) \otimes 
{\bar \partial}_\beta A_2 + \partial_\alpha A_1 \delta^{\alpha \beta} \otimes
{\hat \delta}_D ({\bar \partial}_\beta A_2) \right)~.
\end{aligned}
\end{equation*}
Recall that $\delta^{\alpha\beta}=[z^\alpha\stackrel{\star}{,}\bz^\beta]$ is invariant under twisted diffeomorphisms\footnote{We showed this explicitly in the context of isometries.}, and therefore the derivatives in front of the functions are {\em not} transformed. Instead, one always has to consider $\delta_D$ to be in a fixed representation. Here, this representation is the one acting on functions and therefore it acts trivially on tensor indices; the right transformation law follows thus trivially:
\begin{equation}
\mu(\Delta({\hat \delta}_D ) \CF^{-1}_L A_1 \otimes A_2)\ =\ \hat{\delta}_D(A_1\star A_2)~.
\end{equation}
A different point of view was advocated in \cite{Chaichian:2006we} in the context of noncommutative gauge theories.

\section{Towards fuzzy general relativity}

In this section, we give the construction of the twisted diffeomorphism invariant Einstein-Hilbert action on the fuzzy sphere. This construction is not unique, and it seems that there are several different alternatives. To find the interpretation of these choices as well as to understand the relation with the other theories of gravitation proposed on the fuzzy sphere \cite{Nair:2001kr} will be left to future work. We here consider what seems to be the most direct approach to the problem.

\subsection{The Levi-Civita connection on the fuzzy sphere}

As shown in appendix A, the star product on the fuzzy sphere does not allow for the definition of a square root operation. For this reason, we follow \cite{Aschieri:2005yw} and define a metric from vielbeins:
\begin{equation}
g_{ij}\ =\ \tfrac{1}{2}\left(E_i^m\star E_j^n+E_j^m\star E^n_i\right)\delta_{mn}~.
\end{equation}
The factor $\sqrt{|g|}$ which we inserted to turn tensor densities into honest tensors is then replaced by the star-determinant \cite{Aschieri:2005yw} of the matrix $E$:
\begin{equation}
E^\star\ :=\ \det{}_\star E^m_i\ :=\ \tfrac{1}{4!}\eps_{m_1..m_3}\eps^{i_1\ldots i_3}E^{m_1}_{i_1}\star\ldots \star
E^{m_3}_{i_3}~.
\end{equation}
From here, the definition of the Levi-Civita connection in the noncommutative case follows closely along the lines of the definition of the gauge connection in the noncommutative case. First, we define 
\begin{equation}\label{fuzzyLeviCivita}
\breve{\Gamma}_{ij}^k\ :=\ \tfrac{1}{2}\left(\der{x^i}g_{jl}+\der{x^j}g_{il}-\der{x^l}g_{ij}\right)\star g^{kl}_\star~,
\end{equation}
where $g^{kl}_\star$ is the {\em right-star-inverse} of $g_{ik}$, i.e.\
\begin{equation}
g_{ik}\star g^{kl}\ =\ \delta^l_i~.
\end{equation}
We also followed in our conventions \cite{Aschieri:2005yw}, and put $g^{kl}_\star$ on the right, as we defined it to be a right-inverse. Similarly to the discussion in the commutative situation, we introduce the following modified Levi-Civita connection:
\begin{equation}
\Gamma_{ij}^k\ =\ E^\star\eps_{imn}\star x^m \star\breve{\Gamma}_{rj}^k\star g^{nr}~.
\end{equation}
As in the case of gauge theories, we have to translate the condition $x^i\breve{\Gamma}_{ij}^k=0$ into the noncommutative setting. However, there is a strong difference due to the fact that the action of the gauge group, i.e.\ the group of diffeomorphisms, is twisted, and therefore its action is the same as in the commutative case. For this reason, we can actually impose the same condition on the Christoffel symbols as before: $x^i\Gamma_{ij}^k=0$, or, using the embedding map $\phi_\CA$:
\begin{equation}\label{ChristoffelOperatorCondition}
\phi_\CA(x^i)\phi_\CA(\breve{\Gamma}_{ij}^k)\ =\ 0~.
\end{equation}
One can rewrite this condition using only star products, as one can straightforwardly verify:
\begin{equation}
x^i\star\breve{\Gamma}_{ij}^k+\breve{\Gamma}_{ij}^k\star x^i\ =\ 2\dpar^i\breve{\Gamma}_{ij}^k~.
\end{equation}
It is clear that the physical meaning of this condition has to be studied further. Also, it seems that this condition is not unique.

The na\"ive noncommutative analog to the gauge condition would have read as 
\begin{equation}\label{ChristoffelStarProductCondition}
x^i\star \Gamma_{ij}^k+\Gamma_{ij}^k\star x^i+\frac{1}{L}\Gamma_{in}^k\star \Gamma_{ij}^n\ =\ 0~,
\end{equation}
for which diffeomorphism invariance is not guaranteed. The observation that the condition used to reduce tensor calculus from $\FR^3$ to $\CPP^1_L$ is the same both in the commutative and the fuzzy case, leads to the conclusion that also in gauge theories, one can use the condition $x^iA_i=0$ in the fuzzy case, if one twists the action of the gauge group.

\subsection{Fuzzy Einstein-Hilbert action}

By now, we have all the ingredients to write down a fuzzy version of the Einstein-Hilbert action, i.e.\ an action, which is invariant under twisted fuzzy diffeomorphisms. 

We first introduce the curvature tensor
\begin{equation}
R^k{}_{lij}\ :=\ \CL_i\Gamma^k_{jl}-\CL_j\Gamma^k_{il}+\Gamma^k_{in}\star\Gamma^n_{jl}-\Gamma^k_{jn}\star\Gamma^n_{il}
-\di E^\star\eps_{ijm}\star\Gamma_{nl}^k\star g^{mn}~,
\end{equation}
from which the Ricci tensor and the curvature scalar are calculated by:
\begin{equation}
R_{mn}\ :=\ R^i{}_{min}\eand R\ :=\ R_{mn}\star g^{mn}~.
\end{equation}
The action then reads as
\begin{equation}\label{EHaction}
S=\int\dd^3 x (E^\star \star R + \mbox{c.c.})\delta(x^ax^a-1)~.
\end{equation}
The $\delta$-distribution inside the integral is the commutative one and therefore well defined. Furthermore, it is only radius dependent, and thus invariant under the truncated fuzzy diffeomorphisms, which we are considering. 

To derive the equations of motion, one writes the action in terms of the vielbeins and varies with respect to them. Using the cyclicity of the star product, one can then move the variation of the fields either to the very right or the very left and collect all the terms. Alternatively, one can make an ansatz for the vielbeins and calculate the explicit expansion of all the star products. This is completely analogous to the discussion in \cite{Aschieri:2005yw}.

\subsection{Applications and open questions}

We showed that there are several diffeomorphism-invariant conditions one can write down to reduce the fuzzy tensor calculus on $\FR^3$ to that on $\CPP^1_L$. It would be nice to find the physical meaning of the various conditions. Furthermore, it is also clear that the fields which are put to zero by these conditions could have also been used to construct additional diffeomorphism invariant terms in the Lagrangian. The action we gave in \eqref{EHaction} is certainly only one example of a whole class of fuzzy generalizations of the Einstein-Hilbert action, which are invariant unter the twisted fuzzy diffeomorphisms. Also here, it would be desirable to gain a better understanding of the physics behind the various possible actions.

Though being probably not the most general one, the setup we wrote down is in principle suited for performing numerical studies. For realistic models, one might, however, want to combine the fuzzy sphere with the space $\FR^{1,1}$ to have a four-dimensional spacetime with a metric of Minkowski signature and commutativity in the time direction. In this setting, it would also be interesting to study the existence of fuzzy generalizations of e.g.\ the Schwarzschild solution.

Note that a complete theory of gravity must also allow for topology changes. In performing the decompactification limit as discussed in section 2.3, one can e.g.\ turn the fuzzy sphere into flat $\FR^2_\theta$; the algebra of matrices isomorphic to the algebra of functions, however, will then become an infinite dimensional one. Similarly, a change in the matrix algebra will be obtained by deforming the sphere into a torus, see also \cite{Arnlind:2006ux}. Unfortunately, there is no dynamical way of performing these topology changes in the framework we gave.

Although we studied gravity to give a straightforward example for symmetries which require a twist in the fuzzy setting, it seems that the most important application for the twist introduced in this paper will lie in the application to ordinary fuzzy field theory, as e.g.\ to gauge theories on fuzzy spaces.

\section*{Acknowledgements}

It is a pleasure to thank the participants of the ``Fuzzy Physics and Noncommutative Geometry workshop'' held in Dublin Institute for Advanced Studies 6-20 June 2006 for various fruitful discussions.  In particular, we would like to thank A.P.~Balachandran and D.~O'Connor for critical comments and discussions as well as S.~Murray for discussions and comments on the manuscript. The work of  S.K. is supported by the Irish Research Council for Science Engineering and Technology (IRCSET). The major part of this work was done while C.S. was a postdoctoral fellow within the Marie Curie training network ``Quantum Spaces and Noncommutative Geometry''.

\appendices

\subsection{The star products on $\CPP^n$}

As stated in section 2.2, all the star products are induced from the operator product on an appropriate Fock space via the projector $\hat{\rho}$. The most common such projector is constructed from the truncated coherent states
\begin{equation}
|z,L\rangle\ :=\ \frac{1}{\sqrt{L!}}\left(z_\alpha a_\alpha^\dagger\right)^L|0\rangle~
\end{equation}
as
\begin{equation}\label{Vorosprojector}
\hat{\rho}\ :=\ |z,L\rangle\langle z,L|~,
\end{equation}
and as one easily verifies, this operator defines a map $\tr(\hat{\rho}\,\cdot\,):\hat{\CA}_L\rightarrow \CA_L^\star$:
\begin{equation}\label{mapCoherentState}
\hat{a}^\dagger_{\alpha_1}\ldots \hat{a}^\dagger_{\alpha_L}|0\rangle\langle 0|\hat{a}_{\beta_1}\ldots \hat{a}_{\beta_L}~~~\ \mapsto\ ~~~
\bz^{\alpha_1}\ldots \bz^{\alpha_L}z^{\beta_1}\ldots z^{\beta_L}~.
\end{equation}
As shown in \cite{Balachandran:2001dd}, the projector \eqref{Vorosprojector} yields the star product \eqref{defStarProduct}, which is called the {\em coherent state star product} for obvious reasons. We can now combine the coordinates $z^\alpha,\bz^\alpha$ to
\begin{equation}\label{mapxtoz}
x^a\ :=\ \bz^\alpha \frac{\lambda^a_{\alpha\beta}}{2}z^\beta~,
\end{equation}
where $\lambda^a_{\alpha\beta}$ are the Gell-Mann matrices of $\sSU(n+1)$ and $a=1,\ldots ,(n+1)^2-1$. This contraction amounts to performing the projection in the generalized Hopf fibration \eqref{generalizedHopf} from $S^{2n+1}$ with coordinates $z^\alpha,\bz^\alpha$ down to $\CPP^n$ with coordinates $x^a$ embedded in $\FR^{(n+1)^2-1}$. Thus, polynomials in these coordinates describe functions on $\CPP^n$. To obtain functions on the fuzzy $\CPP^n_L$, we pair off the open indices in the monomials $z^{\alpha_1}\ldots z^{\alpha_L}\bz^{\beta_1}\ldots \bz^{\beta_L}$ with $\frac{\lambda^a_{\gamma\delta}}{2}$ or $\lambda^0_{\gamma\delta}:=\delta_{\gamma\delta}$. Together with $\bz^\alpha z^\alpha=1$, this turns the monomials of degree $2L$ in $z^\alpha,\bz^\alpha$ into monomials which are at most of degree $L$ in $x^a$. These monomials can be used as the basis of the space of functions $\CA_L^\star$. The star product in the new coordinates reads as
\begin{equation*}
(f\star g)(x)\ :=\ \mu\left[\left(\sum_{j=0}^L\frac{(L-j)!}{L!j!}K^{a_1b_1}\ldots K^{a_jb_j}
\der{x^{a_1}}\ldots \der{x^{a_j}}\otimes\der{x^{b_1}}\ldots \der{x^{b_j}}\right)
(f(x)\otimes g(x))\right]
\end{equation*}
with
\begin{equation}
K^{ab}\ =\ \frac{1}{n}\delta^{ab}+\tfrac{1}{\sqrt{2}}(d^{ab}{}_c+\di f^{ab}{}_c)x^c-x^a x^b~.
\end{equation}
There is a number of other star products corresponding to different projectors $\hat{\rho}(p)$ and described by different $K^{ab}$s. Common to all these star products is the antisymmetric part of $K^{ab}$. 

For defining a pseudo-Drinfeld twist, the star product in terms of the real projector coordinates $x^a$ is not useful since it neither simplifies as nicely as \eqref{defStarProduct} nor does it offer a natural way of assigning $K^{ab}$ to the first and second slots of $\der{x^a}\otimes \der{x^b}$. However, once we have defined our pseudo-Drinfeld using the coordinates $z^\alpha,\bz^\alpha$, it evidently carries over to $\FR^{(n+1)^2-1}$, as the map \eqref{mapxtoz} between the coordinate systems is essentially bijective.

Note that the coherent state star product does not allow for a star square root. As a counterexample, consider the case $L=1$ on the fuzzy sphere and let us try to define the square root of $x^1$. A general function reads as
\begin{equation}
f(x)\ =\ a_0+a_1 x^1+a_2 x^2+a_3 x^3~,
\end{equation}
where $a_0,\ldots ,a_3\in\FR$ and its square is
\begin{equation}
f(x)\star f(x)\ =\ a_0^2+a_1^2+a_2^2+a_3^2+2a_0(a_1 x^1+a_2 x^2+a_3 x^3)~.
\end{equation}
There is evidently no way of choosing $a_0,\ldots ,a_3$ to obtain $f(x)\star f(x)=x^1$.

\subsection{Reduction of the star product}

In this section, we prove that the star product \eqref{defStarProduct} is equivalent to \eqref{StarProductReduced} when acting on functions which are elements of $\CA_L^\star$. Although this equivalence is rather obvious from the map \eqref{mapCoherentState}, let us restrict our attention here to formula \eqref{defStarProduct}. Consider first the identity
\begin{equation}
\sum_{l=0}^L(-1)^l \binom{L}{l}\ =\ 0~,
\end{equation}
which can easily be proven, e.g.\ recursively, using identities for the binomial coefficients. Pulling a factor of $L!$ out of the sum, we obtain furthermore
\begin{equation}\label{identAppB}
\sum_{l=0}^L\frac{(-1)^l}{l!(L-l)!}\ =\ 0\eand
\sum_{l=n}^L\frac{(-1)^{l-n}}{(l-n)!(L-l)!}\ =\ \delta_{n,L}~,
\end{equation}
where the second identity is derived from the first one for $n<L$; the case $n=L$ follows trivially.

The terms in the definition of the star product \eqref{defStarProduct} are all of the form
\begin{equation*}
\left(z^{\alpha_1}\ldots z^{\alpha_m}\der{z^{\alpha_1}}\ldots \der{z^{\alpha_m}}\der{z^{\gamma_{1}}}\ldots \der{z^{\gamma_n}}f\right)
\left(\bz^{\beta_1}\ldots \bz^{\beta_m}\der{\bz^{\beta_1}}\ldots \der{\bz^{\beta_m}}\der{\bz^{\gamma_{1}}}\ldots \der{\bz^{\gamma_n}}g\right)~.
\end{equation*}
After inverting the order of the first $m$ derivatives, $m$ Euler operators naturally appear, and for $f,g\in \CA^\star_L$, the above expression is equal to
\begin{equation*}
(L-n)\ldots (L-n-m+1)\left(\der{z^{\gamma_{1}}}\ldots \der{z^{\gamma_n}}f\right)(L-n)\ldots (L-n-m+1)
\left(\der{\bz^{\gamma_{1}}}\ldots \der{\bz^{\gamma_n}}g\right)~.
\end{equation*}
All the terms in the star product proportional to $\left(\der{z^{\gamma_{1}}}\ldots \der{z^{\gamma_n}}f\right)\left(\der{\bz^{\gamma_{1}}}\ldots \der{\bz^{\gamma_n}}g\right)$
are therefore given by 
\begin{equation}
\sum_{l=n}^L\frac{(L-l)!}{L!l!}\binom{l}{n}(-1)^{l-n}\left(\frac{(L-n)!}{(L-n-(l-n))!}\right)^2\ =\ 
\frac{(L-n)!}{L!n!}\delta_{n,L}~,
\end{equation}
where we used the last identity in \eqref{identAppB}. Thus, the star product reduces to the term with $n=L$:
\begin{equation}
(f\star g)(z^\alpha,\bz^\beta)\ :=\ \left(
\frac{1}{L!}\der{z^{\alpha_1}}\ldots \der{z^{\alpha_L}}f\right)\left(\frac{1}{L!}\der{\bz^{\alpha_1}}\ldots \der{\bz^{\alpha_L}}g\right)~,
\end{equation}
as stated in \eqref{StarProductReduced}. Note, however, that this derivation only holds for finite $L$.



\end{document}